\documentstyle[12pt,a4]{article}
\input epsf
\input epsfig.sty

\input{tcilatex}
\begin{document}

\setcounter{page}{0} \topmargin 0pt \oddsidemargin 5mm \renewcommand{%
\thefootnote}{\fnsymbol{footnote}} \newpage \setcounter{page}{0} 
\begin{titlepage}
\begin{flushright}
Berlin Sfb288 Preprint  \\
hep-th/9803005\\
\end{flushright}
\vspace{0.5cm}
\begin{center}
{\Large {\bf Thermodynamic Bethe Ansatz \\ [2mm] with Haldane Statistics} }

\vspace{1.8cm}
 {\large A.G. Bytsko$^{*,**} \; $ \hbox{and}   $\;$  A. Fring$^{*}$ }

\vspace{0.5cm}
{\em $*$ Institut f\"ur Theoretische Physik,
Freie Universit\"at Berlin\\ 
Arnimallee 14, D-14195 Berlin, Germany\\ [2mm]
$**$ Steklov Mathematical Institute\\
Fontanka 27, St.Petersburg, 191011, Russia }
\end{center}
\vspace{1.2cm}
 
\renewcommand{\thefootnote}{\arabic{footnote}}
\setcounter{footnote}{0}

\begin{abstract}

We derive the thermodynamic Bethe ansatz equation for the situation in
which the statistical interaction of a multi-particle system is
governed by Haldane statistics. We formulate a macroscopical
equivalence principle for such systems. Particular CDD-ambiguities
play a distinguished role in compensating the ambiguity in the 
exclusion statistics. We derive Y-systems related to generalized statistics.
We discuss several fermionic, bosonic and anyonic versions of affine
Toda field theories and Calogero-Sutherland type models
in the context of generalized statistics.

\bigskip

\end{abstract}
\vspace{.3cm}
\centerline{March 1998}
\vfill{ \hspace*{-9mm}
\begin{tabular}{l}
\rule{6 cm}{0.05 mm}\\
Bytsko@physik.fu-berlin.de ,\ \ Bytsko@pdmi.ras.ru \\
Fring@physik.fu-berlin.de
\end{tabular}}
\end{titlepage}
\newpage

\section{Introduction}

The Bethe ansatz is a technique which is based upon the quantum mechanical
description of a many particle system by a wave function. The interaction of
the individual particles is assumed to be relativistic, short-range and
characterized by a factorizable scattering matrix. The boundary condition
for the many particle wave function leads to what is commonly referred to as
the Bethe ansatz equation, which provides the quantization condition for
possible momenta of this system. Taking the thermodynamic limit of this
equation, that is taking the size of the quantizing system to infinity,
leads to the so-called thermodynamic Bethe ansatz (TBA). The origins of this
analysis trace back to the seminal papers by Yang and Yang \cite{Yang} and
the technique has been refined and applied to different situations in
numerous works [2-6] thereafter. The TBA
constitutes an interface between massive integrable models and conformal
field theories. One may extract different types of information from it,
where the ultraviolet behaviour, i.e. ultimately the central charge of the
conformal field theory, is the most accessible.

In the derivation of the TBA-equation the underlying statistical interaction 
(also called exclusion statistics) is usually taken to be
either of bosonic or fermionic type. Seven years ago Haldane \cite{Haldane}
proposed a generalized statistics based upon a generalization of Pauli's
exclusion principle. This type of statistics (anyonic) has many important
applications, in particular in the description of the fractional quantum
Hall effect \cite{FQHE}. The main purpose of this manuscript is to 
implement systematically the Haldane statistics into the analysis of
the thermodynamic Bethe ansatz.
Hitherto, attempts in this direction \cite{Wu,Poly} were 
mainly based on the consideration
of  particular statistical interaction which only involves one 
species, like the Calogero-Sutherland model \cite{Calogero}.  Our
approach will cover a general choice of statistical interaction 
described by some, in general non-diagonal, matrix $g_{ij}$.
We put particular emphasis on the ultraviolet region, which 
corresponds to the high temperature regime. 
We also continue our investigation started in \cite{ABAF} and
clarify the role of the anyonic S-matrix (\ref{anS}) in the
context of the TBA.

We formulate a macroscopical equivalence principle in the sense that the
macroscopical nature of a multi-particle system is only governed by the
combination (\ref{inva}) of the dynamical- and statistical interactions.
This means in particular that two multi-particle systems differing on the
microscopical level, i.e. in the S-matrix, may be made macroscopically
equivalent by tuning the statistical interaction.

Our manuscript is organized as follows: In section 2 we recall the
derivation of several thermodynamic quantities from Haldane statistics. In
section 3 we derive the thermodynamic Bethe ansatz equation for a
multi-particle system in which the statistical interaction is governed by
Haldane statistics. In section 4 we argue that certain multi-particle
systems may be transformed into macroscopically equivalent systems by tuning
the statistical- and the dynamical interaction. In section 5 we demonstrate
that certain scattering matrices, leading to equivalent thermodynamical
systems differ only by CDD-ambiguities and comment on the ambiguity in
choosing a particular statistical interaction. In section 6 we discuss the
ultraviolet limit of the generalized TBA-equation. In
section 7 we derive Y-systems related to generalized statistics. In section
8 we illustrate our general statements by some explicit examples. Our
conclusions are stated in section 9.

\section{Thermodynamics from Haldane Statistics}

The object of our consideration is a multi-particle system containing $l$
different species confined to a finite region of size $L$. We denote by $%
n_{i}$ the number of particles, by $N_{i}$ the dimension of the 
Fock-space related to the
species $i$ and by $d_{i}$ the number of available states
(holes) before the $n_i$-th  particle has been added to the system. 
When treating bosons the number of available states naturally
equals the total dimension of the Fock space, i.e. $N_{i}=d_{i}$, whereas
when treating fermions there will be restrictions due to Pauli's exclusion
principle and one has $N_{i}=d_{i}+n_{i}-1$. With these relations in mind
the total dimension of the Hilbert space may be written for both cases as 
\begin{equation}
W=\prod\limits_{i=1}^{l}\frac{(d_{i}+n_{i}-1)!}{n_{i}!(d_{i}-1)!}\quad .
\label{Wahr}
\end{equation}
Conventionally one employs in (\ref{Wahr}) the Fock-space dimension rather
than the number of available states  before the $n_i$-th  particle has 
been added to the
system. However, besides being a unified
formulation, equation (\ref{Wahr}) has the virtue that it allows for a
generalization to Haldane statistics.

By introducing a statistical interaction $g_{ij},$ Haldane \cite{Haldane}
proposed the following generalized Pauli exclusion principle 
\begin{equation}
\frac{\Delta d_{i}}{\Delta n_{j}}=-g_{ij}\quad .  \label{Pauli}
\end{equation}
Relation (\ref{Pauli}) means that the number of available states should be
regarded as a function of the particles present inside the system. In this
proposal a statistical interaction between different particle species is
conceivable. In the bosonic case there will be no restriction such that $%
g_{ij}=0$, whereas in the fermionic case the number of available states
reduces by one if a particle is added to the system, hence one chooses $%
g_{ij}=\delta _{ij}.$ One assumes \cite{Haldane,Wu}, that the total
dimension of the Hilbert space is still given by (\ref{Wahr}), where the
quantities involved are related to each other by (\ref{Pauli}).

We now want to analyze the multi-particle system in its thermodynamic limit,
that is we let the size of the confining region approach infinity, $%
L\rightarrow \infty $. It is then a common assumption that the ratio of the
particle (hole or state) numbers over the system size $L$ remains finite.
For the fraction of particles (holes or states) of species $i$ with
rapidities between $\theta -\Delta \theta /2$ and $\theta +\Delta \theta /2$
it is convenient to introduce densities 
\begin{eqnarray}
\Delta N_{i} &=&\rho _{i}(\theta )\,\Delta \theta \,L  \label{D1} \\
\Delta n_{i} &=&\rho _{i}^{r}(\theta )\,\Delta \theta \,L  \label{D2} \\
\Delta d_{i} &=&\rho _{i}^{h}(\theta )\,\Delta \theta \,L\quad .  \label{D3}
\end{eqnarray}
The rapidity $\theta $ parameterizes as usual the two-momentum $\vec{p}%
=m\left( \cosh \theta ,\sinh \theta \right) $. Integration of the
generalized Pauli exclusion principle (\ref{Pauli}) then yields a relation
between the different types of densities 
\begin{equation}
\rho _{i}(\theta )\,=\rho _{i}^{h}(\theta
)\,+\sum\limits_{j=1}^{l}g_{ij}\rho _{j}^{r}(\theta )\,\quad .
\label{Dichte}
\end{equation}
The constant of integration has been identified with the Fock space
dimension. The reason for this identification is based on fact that in this
way one recovers for $g_{ij}=0$ and $g_{ij}=\delta _{ij\text{ }}$ the usual
bosonic and fermionic relations, respectively. Notice that in the finite
case the constant is in general slightly different from $N_{i}$, for
instance for fermionic statistics it has to be chosen as $N_{i}+1$ in order
to recover the relation $N_{i}=d_{i}+n_{i}-1$. However, in the thermodynamic
limit this difference is negligible.

Let us remark, that equation (\ref{Dichte}) suggests from a physical point
of view that $g_{ij}$ should be non-negative. However, if there are
additional symmetries then this requirement can be weakened. For instance
for two conjugate particles, say $i$ and $\bar{\imath}$ (see section 8 for
examples), one naturally assumes that $\rho _{i}^{r}(\theta )=\rho _{\bar{%
\imath}}^{r}(\theta )$, such that only the combination $g_{ji}+g_{j\bar{%
\imath}\text{ }}$ has to be taken non-negative.

We are now in the position to construct further thermodynamic quantities.
First of all we may sum up all contributions from occupied states in order
to obtain the total energy 
\begin{equation}
E\left[ \rho ^{r}\,\right] =L\sum\limits_{i=1}^{l}\int\limits_{-\infty
}^{\infty }d\theta \rho _{i}^{r}(\theta )m_{i}\cosh  \theta 
\quad .  \label{ener}
\end{equation}
Furthermore, we obtain from (\ref{Wahr}), upon using Stirling's formula $\ln
n!\approx n\ln n$, (\ref{D1})-(\ref{D3}) and (\ref{Dichte}), the
entropy $S=k\ln W$ as a functional of the particle- and Fock-space density 
\begin{equation}
S\left[ \rho ,\rho ^{r}\,\right]
=kL\sum\limits_{i=1}^{l}\int\limits_{-\infty }^{\infty }d\theta \left[ (\rho
_{i}-g_{ij}\rho _{j}^{r})\ln \left( \frac{\rho _{i}+h_{ij}\rho _{j}^{r}}{%
\rho _{i}-g_{ij}\rho _{j}^{r}}\right) +\rho _{i}^{r}\ln \left( \frac{\rho
_{i}+h_{ij}\rho _{j}^{r}}{\rho _{i}^{r}}\right) \right] \quad .
\label{entro}
\end{equation}
We introduced here $h_{ij}=\delta _{ij}-g_{ij}$ and use the sum convention
(sum over j) to avoid bulky expressions. $k$ is Boltzmann's constant.
According to the fundamental postulates of thermodynamics the equilibrium
state of a system is found by minimizing the free energy $F$. Hence, keeping
the temperature constant we may obtain the equilibrium condition by
minimizing $F\left[ \rho ,\rho ^{r}\,\right] =E\left[ \rho ^{r}\,\right]
-TS\left[ \rho ,\rho ^{r}\,\right] $ with respect to $\rho ^{r}$. The
equilibrium condition reads 
\begin{equation}
\frac{\delta F}{\delta \rho _{i}^{r}}=\frac{\delta E}{\delta \rho _{i}^{r}}-T%
\frac{\delta S}{\delta \rho _{i}^{r}}-T\sum\limits_{j=1}^{l}\frac{\delta S}{%
\delta \rho _{j}}\frac{\delta \rho _{j}}{\delta \rho _{i}^{r}}=0\quad .
\label{equ}
\end{equation}
So far we did not provide any information about the admissible momenta in
the system, which are restricted by the boundary conditions.

\section{Thermal Equilibrium with Boundaries}

Boundary conditions may be accounted for by the Bethe ansatz equations.
Recall that the Bethe ansatz equation is simply the equation which results
from taking a particle in the multi-particle wave function on a trip through
the whole system \cite{Yang}. The particle will scatter with all other
particles in the system, described by a factorizable S-matrix, such that 
\begin{equation}
\exp (iLm_{i}\sinh \theta _{i} )\prod\limits_{j\neq i}^{l}S_{ij}(\theta
_{i}-\theta _{j})=1\quad ,  \label{BA}
\end{equation}
has to hold for consistency \cite{Yang}. To simplify notations we may assume
here that the scattering matrix is diagonal, such that the subscripts only
label particle species. This set of transcendental equations determines  
which rapidities are admissible in the system due to the quantization 
as a result of restricting the size of the system. Taking the logarithmic 
derivative of the Bethe ansatz equation (\ref{BA}) and employing densities 
as in (\ref{D1}) and (\ref{D2}) one obtains (see for instance 
\cite{TBAZam}, \cite{Roland} or \cite{TBAKM} for more details) 
\begin{equation}
\frac{1}{2\pi }m_{i}\cosh \theta +\sum\limits_{j=1}^{l}\left( \varphi
_{ij}*\rho _{j}^{r}\right) (\theta )=\rho _{i}(\theta )\quad .  \label{Bou}
\end{equation}
Here we introduced as usual the notation $\varphi _{ij}(\theta )=-i\frac{d}{%
d\theta }\ln S_{ij}(\theta )$ and denote the convolution by $\left(
f*g\right) (\theta )$ $:=1/(2\pi )\int d\theta ^{\prime }f(\theta -\theta
^{\prime })g(\theta ^{\prime })$. The equilibrium condition (\ref{equ})
together with (\ref{Bou}) yields the desired thermodynamic Bethe ansatz
equation for a system in which statistical interaction is governed by
Haldane statistics 
\begin{equation}
\frac{1}{kT}m_{i}\cosh \theta =\ln (1+x_{i}\left( \theta \right)
)+\sum\limits_{j=1}^{l}\left( \Phi _{ij}*\ln (1+x_{j}^{-1})\right) (\theta
)\quad .  \label{BFTBA}
\end{equation}
We use here the abbreviations $x_{i}\left( \theta \right) :=\rho
_{i}^{h}\left( \theta \right) /\rho _{i}^{r}\left( \theta \right) $ and 
\begin{equation}
\Phi _{ij}(\theta ):=\varphi _{ij}\left( \theta \right) -2\pi g_{ij}\delta
\left( \theta \right) \quad .  \label{inva}
\end{equation}
In the derivation of (\ref{BFTBA}) we assumed that $g_{ij}=g_{ji}$.
We assume that $x_{i}\left( \theta \right) $ is symmetric\footnote{%
Taking into account that $\Phi _{ij\text{ }}(\theta )$ is symmetric in $%
\theta $ (due to (\ref{uni})), this is equivalent to the assumption that (%
\ref{BFTBA}) may be solved iteratively (see e.g. \cite{Korepin}).} in the
rapidity throughout the manuscript. In general one is only able to solve
equations (\ref{BFTBA}) numerically as we demonstrate in section 8. In the
formulation of the TBA-equation of bosonic or fermionic type, it is common
to introduce here as an additional quantity the so-called pseudo-energies $%
\varepsilon _{i}(\theta ).$ In general one may employ 
\begin{equation}
\ln (1+x_{i}(\theta ))-\sum\limits_{j=1}^{l}g_{ij}\ln (1+x_{j}^{-1}(\theta
))=\varepsilon _{i}(\theta )\quad ,  \label{Id1}
\end{equation}
for this purpose. For the finite case the same relation was obtained by Wu 
\cite{Wu}, there however, the quantity $x_{i}(\theta )$ has a slightly
different meaning. Clearly it is not possible to give a general solution of (%
\ref{Id1}). However, for the bosonic and fermionic case it is solved easily,
we obtain $x_{i}(\theta )=\exp (\varepsilon _{i}(\theta ))-1$ and $%
x_{i}(\theta )=\exp (\varepsilon _{i}(\theta ))$. Substitution of these
solutions into equations (\ref{BFTBA}) turns them into the well-known
TBA-equations of bosonic and fermionic type \cite{TBAZam}, respectively.
With (\ref{Dichte}) the ratios of the related particle- and Fock-space
densities 
\begin{equation}
\frac{\rho _{i}^{r}(\theta )}{\rho _{i}(\theta )}=\frac{1}{\exp (\varepsilon
_{i}(\theta ))\mp 1}\qquad  \label{BEFD}
\end{equation}
become the usual Bose-Einstein (upper sign) and Fermi-Dirac (lower sign)
distributions. From our point of view it does not seems to be necessary to
introduce pseudo-energies and one should rather view $x_{i}\left( \theta
\right) =\rho _{i}^{h}\left( \theta \right) /\rho _{i}^{r}\left( \theta
\right) $ as a more fundamental entity. In addition one avoids the problem
of solving (\ref{Id1}). From a numerical point of view however, it appears
sometimes useful to formulate the TBA-equation in different variables.

We now substitute back the equilibrium condition into the expression for the
free energy and obtain together with the general expressions for the total
energy (\ref{ener}) and the entropy (\ref{entro}), the generalized
thermodynamic Bethe ansatz equation (\ref{BFTBA}) and (\ref{Bou}) 
\begin{equation}
F(T)=-\frac{LkT}{2\pi }\sum\limits_{i=1}^{l}\int\limits_{-\infty }^{\infty
}d\theta m_{i}\cosh \theta \ln \left( 1+x_{i}^{-1}(\theta )\right) \quad .
\label{Free}
\end{equation}
The relation between the free energy and the finite size scaling function is
well-known to be $c(T)=-6F(T)/(\pi LT^{2})$ \cite{Cardy}. As usual we now
identify the temperature with the inverse of one radial size of a torus%
\footnote{%
For more details on the physical picture see for instance \cite
{TBAZam,Roland,TBAKM,Cardy}.} $T=1/r$ and choose now Boltzmann's constant to
be one. Then 
\begin{equation}
c(r)=\frac{6r}{\pi ^{2}}\sum\limits_{i=1}^{l}m_{i}\int\limits_{0}^{\infty
}d\theta \cosh \theta \ln \left( 1+x_{i}^{-1}(\theta )\right) \quad .
\label{c(r)}
\end{equation}
Once more for $x_{i}(\theta )=\exp (\varepsilon _{i}(\theta ))-1$ and $%
x_{i}(\theta )=\exp (\varepsilon _{i}(\theta ))$ we recover the well-known
expressions for the scaling functions of bosonic and fermionic statistics. In
the ultraviolet limit the scaling function becomes the effective central
charge of a conformal field theory \cite{Cardy}, i.e. $\lim_{r\rightarrow
0}c(r)=c_{eff}=c-24h^{\prime }$. Here $c$ is the usual conformal anomaly and 
$h^{\prime }$ denotes the lowest conformal dimension \cite{BPZ}.

To summarize: For a given statistical interaction (\ref{Pauli}) and
dynamical interaction described by a factorizable scattering matrix, we may
solve in principle the generalized thermodynamic Bethe ansatz equation (\ref
{BFTBA}) for $x_{i}(\theta )$. This solution together with the knowledge of
the mass spectrum of the theory, allows (up to a one dimensional integral
which may always be carried out by simple numerics) the calculation of the
entire scaling function.

\section{Equivalent Multi-Particle Systems}

{}From a thermodynamic point of view, systems which have the same expressions
for the free energy show the same behaviour. Hence multi-particle systems
which possess the same scaling function over the entire range of the scaling
parameter $r$ are to be considered as equivalent. This implies that two
multi-particle systems are equivalent, if the mass spectra and the
quantities $\Phi _{ij}$ are identical. As a consequence of this we may
achieve that two systems equal each other from a macroscopical point of
view, despite the fact that they involve different scattering matrices
describing the dynamical particle interaction. The apparent difference can
be compensated by a different choice of the underlying statistical
interaction. Considering for instance the well-known TBA-equations of
bosonic- and fermionic type (see for instance \cite{TBAZam}) involving $%
\varphi _{ij}^{b}(\theta )$ and $\varphi _{ij}^{f}(\theta )$, respectively,
we observe that 
\begin{equation}
\varphi _{ij}^{f}(\theta )=\varphi _{ij}^{b}(\theta )+2\pi \delta
_{ij}\delta (\theta )  \label{lodiff}
\end{equation}
transforms both equations into each other.

The question of whether relations between two scattering matrices leading to
(\ref{lodiff}) are at all conceivable immediately comes to mind. Such
relations emerge in several places. For example in \cite{Rava} the authors
assume 
\begin{equation}
S_{ij}^{\prime }(\theta )=S_{ij}(\theta )\exp (-2\pi i\delta _{ij}\Theta
(\theta ))\,.  \label{SS}
\end{equation}
The function $\Theta (\theta )$ was taken to be the usual step-function, with
the property that $\Theta (0)=1/2$, such that $S_{ij}(\theta )$ and $%
S_{ij}^{\prime }(\theta )$ only differ at the origin of the complex rapidity
plane. This modification was necessary in order to extend the validity of
certain identities ((2.18) in \cite{Rava} leading to eq. (\ref{RAD})
below) involving
the scattering matrices of ADE-affine Toda field theories (see for instance 
\cite{TodaS}), i.e. $S_{ij}(\theta )$, also to the origin. In particular, in
some special cases these identities become equal to the bootstrap equations.
A further motivation to introduce (\ref{SS}) was to derive so-called
Y-systems proposed by Zamolodchikov (last ref. in \cite{TBAZam}) for
ADE-affine Toda field theories. In this case the system which involves $%
S_{ij}(\theta )$ with fermionic statistical interaction is equivalent to the
system involving $S_{ij}^{\prime }(\theta )$ with bosonic statistical
interaction.

As a further example we may also consider a system in which the dynamical
interaction is described by the scattering matrix $S(\theta )=-\exp (-i\pi
\lambda \epsilon (\theta ))$, where $\lambda $ denotes the coupling constant
and $\epsilon (\theta )=\Theta (\theta )-\Theta (-\theta )$, of the
Calogero-Sutherland model \cite{Calogero} with a statistical interaction of
bosonic type, i.e. $g=0$. This system is equivalent to a system with a
constant S-matrix, e.g. $S^{\prime }(\theta )=-1$, and a statistical
interaction of the form $g^{\prime }=\lambda .$

In general we have the following equivalence principle. Multi-particle
systems, involving quantities such that the relation 
\begin{equation}
\varphi _{ij}^{\prime }(\theta )=\varphi _{ij}(\theta )+2\pi \left(
g_{ij}^{\prime }-g_{ij}\right) \delta (\theta )  \label{trans}
\end{equation}
holds and the masses  for the same species are identical, 
are thermodynamically equivalent.

\section{Microscopical Ambiguities}

\subsection{On CDD-Ambiguities}

We now want to elaborate on the question in which sense the scattering
matrices leading to (\ref{trans}) differ. The analysis of analytic
properties of the scattering matrix leads to a set of consistency equations,
which have to be satisfied by any S-matrix related to integrable models in
1+1 dimensions. These equations are so restrictive, that they determine the
S-matrices of a particular model up to what is usually referred to as
CDD-ambiguities \cite{CDD,Mitra}. In this section we argue that certain
S-matrices related to each other in such a way, that they lead to (\ref
{trans}), differ precisely by such ambiguities.

A scattering matrix $S_{ij}(\theta )$ is usually assumed to be a meromorphic
function in the strip $0<\func{Im}\theta <\pi ,-\infty <\func{Re}\theta
<+\infty $. This region is considered as physical, meaning that all
singularities occurring in this sheet acquire a physical interpretation. The
scattering matrix in (\ref{BA}) is usually regarded as the one which results
from an analysis of the so-called bootstrap equations 
\begin{eqnarray}
S_{ij}(\theta )S_{ij}(-\theta ) &=&1  \label{uni} \\
S_{ij}(\theta )S_{\bar{\imath}j}(\theta -i\pi ) &=&1  \label{cross} \\
S_{li}(\theta +i\eta )S_{lj}(\theta +i\eta ^{\prime }) &=&S_{l\bar{k}%
}(\theta )\quad .  \label{boot}
\end{eqnarray}
Equation (\ref{uni}) is a result of unitarity and analytic continuation, (%
\ref{cross}) a consequence of crossing invariance and (\ref{boot}) (strictly
speaking this is the boostrap equation) expresses the factorization property
for the fusing process $i+j\rightarrow \bar{k}$. The so-called fusing angles 
$\eta ,\eta ^{\prime }$ are specific to each model depending on the mass
spectrum. It was found in \cite{Mitra} that the most general solution to
these consistency equations will always be of the form 
\begin{equation}
\prod\limits_{\alpha \in A}\frac{\tanh \frac{1}{2}\left( \theta +\alpha
\right) }{\tanh \frac{1}{2}\left( \theta -\alpha \right) }\quad ,
\label{allgsol}
\end{equation}
where $A$ is a set of complex numbers which characterizes a particular
model. When particles are not self-conjugate one should replace $\tanh$
by $\sinh$ in (\ref{allgsol}). There is however the freedom to multiply 
these expressions with a
so-called CDD-ambiguity \cite{CDD} (also of the form (\ref{allgsol}), but
related to a different set $A^{\prime }$), which satisfies by itself all the
consistency requirements without introducing any additional poles into the
physical sheet. A well-known example for such an ambiguity are for instance
the coupling constant dependent blocks. One seeks solutions of the general
form (\ref{allgsol}) involving two sets $A$ and $A^{\prime }$ such that the
product related to the set $A$ (the so-called minimal S-matrix) already
closes the bootstrap and accounts for the whole particle spectrum
independent of the coupling constant. Then the additional factors related to 
$A^{\prime }$ constitute a CDD-ambiguity depending on the coupling constant $%
\beta $ in such a way, that in the limit $\beta \rightarrow 0$ (in theories
which admit duality also $\beta \rightarrow \infty $) the S-matrix becomes
free, that is one.

As a particular case of two S-matrices related to each other such that they
may lead to an equation of the type (\ref{trans}) we will now consider 
\begin{equation}
S_{ij}^{\prime }(\theta )=S_{ij}(\theta )\exp (-2\pi i\Delta _{ij}^{+}\Theta
(\func{Re}\theta )-2\pi i\Delta _{ij}^{-}\Theta (-\func{Re}\theta ))\quad ,
\label{anS}
\end{equation}
which we discussed in \cite{ABAF}. Similarly as in \cite{Rava} we choose $%
\Theta (0)=1/2$. Here the $\Delta _{ij}^{\pm }$ are related to the
asymptotic phases of the S-matrix $S_{ij}(\theta )$%
\begin{equation}
\lim_{\func{Re}\theta \rightarrow \pm \infty }S_{ij}(\theta )=\exp \left(
2\pi i\Delta _{ij}^{\pm }\right) \,.  \label{asphase}
\end{equation}
The asymptotic phases are well defined, since the limit $\func{Re}\theta
\rightarrow \pm \infty $ of (\ref{allgsol}) does not depend on the imaginary
part of $\theta $. This property is in particular needed to obtain (\ref
{bootph}). The transformation (\ref{anS}) compensates the asymptotic phases
and creates a non-trivial phase at $\theta \rightarrow 0$, i.e. the anyonic
situation. It was argued \cite{KT}, that from a physical point of view,
S-matrices which possess a non-trivial asymptotic phase should be regarded
rather as auxiliary objects. Scattering matrices of the type (\ref{anS})
should be considered as the genuine physical quantities, since they lead to
the correct physical properties, including the exchange statistics.

We shall demonstrate that the system of equations (\ref{uni})-(\ref{boot})
does not have to be altered for the anyonic matrix $S_{ij}^{\prime }(\theta
) $. For this purpose we will first derive some properties concerning the
phases. Taking the limit $\func{Re}\theta $ $\rightarrow \infty $ in (\ref
{uni})-(\ref{boot}) yields immediately several relations between the
asymptotic phases 
\begin{eqnarray}
\Delta _{ij}^{+}+\Delta _{ij}^{-} &=&n  \label{uniph} \\
\Delta _{\bar{\imath}j}^{\pm }+\Delta _{ij}^{\pm } &=&n_{\pm }^{\prime }
\label{crossph} \\
\Delta _{li}^{\pm }+\Delta _{lj}^{\pm } &=&\Delta _{l\bar{k}}^{\pm }+n_{\pm
}^{\prime \prime }\quad ,  \label{bootph}
\end{eqnarray}
with $n,n_{\pm }^{\prime },n_{\pm }^{\prime \prime }$ being some integers.
Using now these relations for the phases together with the fact that $%
S_{ij}(\theta )$ satisfies the consistency equations (\ref{uni})-(\ref{boot}%
), it is straightforward to derive the related equations for the anyonic
scattering matrix $S_{ij}^{\prime }(\theta )$%
\begin{eqnarray}
S_{ij}^{\prime }(\theta )S_{ij}^{\prime }(-\theta ) &=&1  \label{unih} \\
S_{ij}^{\prime }(\theta )S_{\bar{\imath}j}^{\prime }(\theta -i\pi ) &=&1
\label{crossh} \\
S_{li}^{\prime }(\theta +i\eta )S_{lj}^{\prime }(\theta +i\eta ^{\prime })
&=&S_{l\bar{k}}^{\prime }(\theta )\quad .  \label{booth}
\end{eqnarray}
Comparing $S_{ij}(\theta )$ and $S_{ij}^{\prime }(\theta )$, we conclude,
taking (\ref{uniph}) into account, that the additional factor in (\ref{anS})
has altered the behaviour at the imaginary axis of the rapidity plane only
up to a sign. Thus $S_{ij}(\theta )$ and $S_{ij}^{\prime }(\theta )$ differ
by a CDD-factor.

It is straightforward to generalize the previous argument to the case when
the scattering matrix is non-diagonal, such that in addition to (\ref{uni})-(%
\ref{boot}) one also has to satisfy the Yang-Baxter equation as a
consequence of factorization.

\subsection{On statistical Ambiguities}

In this subsection we recall the argument which leads to a particular choice
of the statistical interaction. In general one considers the value of the
scattering matrix at $\theta =0$ in order to deduce the symmetry properties
of the Bethe wave function. Then together with the a priori (or e.g. from a
Lagrangian) knowledge of the nature of the particles one deduces the
statistical interaction. For example for $S_{ii}(0)=-1$ the Bethe wave
function is antisymmetric. If in this case one describes bosons, one is
forced to choose a fermionic statistical interaction. However, this way of
arguing seems somewhat ambiguous as a simple example demonstrates.
Considering for instance a system in which the dynamical interaction is
described by the affine Toda scattering matrix, one usually selects \cite
{TBAKM} fermionic statistics due to the fact that $S_{ij}(0)=(-1)^{\delta
_{ij}}$. In the limit $\beta \rightarrow 0$ (because of strong-weak duality
one may also take $\beta \rightarrow \infty $) the theory becomes free and
one obtains $S_{ij}(\theta )=1$. Using now the same arguments one has to
deduce, from the symmetry of the Bethe wave function and the fact that one
still describes bosons, that the statistical interaction has to be bosonic.
Concerning the scaling function there is no problem here since in both cases
we obtain the same ultraviolet limit. However, there is a change in the
statistics. The apparent paradox is resolved by making use of the
macroscopical equivalence. Using the scattering matrix (\ref{SS}) instead
one has now $S_{ij}(0)=1$. Making then use of the bosonic-fermionic
transformation (\ref{lodiff}) we obtain by means of the same arguments a
unique, that is bosonic, statistical interaction for the entire range of the
coupling constant $\beta $.

Scattering matrices related to each other as in (\ref{anS}) play a
distinguished role in this context since they only differ by a CDD-ambiguity
as argued in the previous section. Considering such expressions the relation
between the corresponding quantities $\varphi _{ij}^{\prime }(\theta )$ and $%
\varphi _{ij}(\theta )$ are fixed. In this case we achieve thermodynamical
equivalence by demanding 
\begin{equation}
g_{ij}^{\prime }-g_{ij}=\Delta _{ij}^{-}-\Delta _{ij}^{+}\quad .
\end{equation}
We may consider a few examples. For instance in the case of affine Toda
field theory we have $\Delta _{ij}^{\pm }=\pm \delta _{ij}/2$ \cite{TBAKM},
such that $g_{ij}^{\prime }-g_{ij}=-\delta _{ij}$, 
which is a transformation from fermionic exclusion statistics to bosonic one. 
In case we only consider the minimal part
of the scattering matrix for ADE-affine Toda field theory we obtain from the
asymptotic behaviour observed in \cite{TBAKM}, i.e. $\Delta _{ij}^{\pm }=\pm
(\delta _{ij}/2-(C^{-1})_{ij})$, that is $g_{ij}^{\prime }-g_{ij}=C^{-1}I$ ,
where $C$ denotes the Cartan matrix and $I$ the incidence matrix of the
related Lie algebra. Assuming now that $g_{ij}$ is fixed by the arguments
presented above, we obtain for the later case an interesting expression for
the statistical interaction $g_{ij}^{\prime }$ in terms of Lie algebraic
quantities. Therefore one may formulate the generalized Pauli principle in
this context in a Lie algebraic form.

\section{The ultraviolet Limit}

One of the interesting quantities which may be extracted from the
thermodynamic Bethe ansatz is the effective central charge of the conformal
field theory when taking the ultraviolet limit \cite{Cardy}, i.e. $%
r\rightarrow 0$ in (\ref{c(r)}). The integral equation (\ref{BFTBA})
simplifies in this case to a set of constant coupled non-linear equations 
\begin{equation}
\ln (1+x_{i})=\sum\limits_{j=1}^{l}\left( N_{ij}+g_{ij}\right) \ln
(1+x_{j}^{-1})\quad ,  \label{constTBA}
\end{equation}
where $N_{ij}=\Delta_{ij}^- - \Delta_{ij}^+ $. In this limit 
one may approximate $rm_{i}\cosh
\theta $ in (\ref{BFTBA}) and (\ref{c(r)}) by $\exp \theta $ $rm_{i}/2$ 
\footnote{%
Of course these approximations rely upon certain assumptions (for details see
for instance \cite{Hrachik,TBAZam,TBAKM}).}. Taking thereafter the
derivative of (\ref{BFTBA}) we obtain, upon the substitution of the result
into (\ref{c(r)}), for the effective central charge 
\begin{equation}
c_{eff}=\frac{6}{\pi ^{2}}\sum\limits_{i=1}^{l}L\left( \frac{1}{1+x_{i}}%
\right) \quad .  \label{ceff}
\end{equation}
Here $L(x)$ $=-\frac{1}{2}\int_{0}^{x}dt\left[ \frac{\ln (1-t)}{t}+\frac{\ln
t}{1-t}\right] =\sum_{n=1}^{\infty }\frac{x^{n}}{n^{2}}+\frac{1}{2}\ln x\ln
(1-x)$ denotes Rogers dilogarithm \cite{Lewin}. In these definitions it is
assumed that $x$ takes its values between 0 and 1, which in turn implies
that all $x_{i}$ in (\ref{ceff}) are to be non-negative. This is in
agreement with the physical interpretation of the $x_{i}$ as ratios of
densities. Once again with $x_{i}=\exp (\varepsilon _{i})-1$ and $x_{i}=\exp
(\varepsilon _{i})$ we recover the well-known expressions for the bosonic
and fermionic type of statistical interaction, respectively.

Obviously the transformation properties discussed in section 4 also 
survive  this limit process. 
As follows directly from (\ref{constTBA}) they read now 
\begin{equation}
N_{ij}^{\prime }=N_{ij}+g_{ij}-g_{ij}^{\prime }\quad .
\end{equation}
Apparently this condition is weaker than (\ref{trans}). It guarantees
equivalence multi-particle systems only at the conformal point. 
The transformation from bosonic to fermionic statistics is compatible with
equations (50) and (51) in \cite{ABAF} for the case b=1, where such
transformations where obtained purely on the conformal level, that is by
means of manipulations of certain characters.

It is instructive to consider a few examples. For instance having the
situation that the difference of the phases of the scattering matrix equals
the negative of the statistical interaction, i.e. $N_{ij}=-g_{ij}$ we always
obtain 
\begin{equation}
c_{eff}=\frac{6}{\pi ^{2}}\sum\limits_{i=1}^{l}L\left( 1\right) =l\quad .
\label{37}
\end{equation}
An example for this situation is to consider a system involving the
scattering matrix related to ADE-affine Toda field theory (recall that in
this case $N_{ij}=-\delta _{ij}$ \cite{TBAKM}) and choose the statistical
interaction to be fermionic.

An interesting structure emerges when considering a system in which the
dynamical interaction is described by a direct product of
Calogero-Sutherland scattering matrices, i.e. 
$S_{ij}(\theta )=-\exp (-i\pi \lambda
_{i}\delta _{ij}\epsilon (\theta ))$, and the statistical interaction is
taken to be of fermionic type. Then (\ref{constTBA}) reduces to 
\begin{equation}
x_{i}^{1+\lambda _{i}}=\left( 1+x_{i}\right) ^{\lambda _{i}}\quad .
\label{ctba}
\end{equation}
The same equation is of course obtained when the statistical interaction is
chosen to be bosonic and the coupling constants shifted by one. In some
cases (\ref{ctba}) is solved easily analytically and we can employ (\ref
{ceff}) to compute the effective central charge. We may decompose the
effective central charge into contributions coming from different choices
for the $\lambda _{i}$: $c_{eff}=\sum_{i=1}^{l}c_{eff}^{i}$. For instance
for $\lambda _{i}=1$ the solution of (\ref{ctba}) is $x_{i}=(1+\sqrt{5})/2$
and for $\lambda _{i}=-1/2$ we obtain $x_{i}=(\sqrt{5}-1)/2$. Then for $%
\lambda _{i}=-1/2$ and $\lambda _{i}=1$ we obtain, with the help of (\ref
{ceff}), 
\begin{equation}
c_{eff}^{i}=\frac{6}{\pi ^{2}}L\left( \frac{2}{1+\sqrt{5}}\right) =\frac{3}{5%
}\quad \quad \text{and\quad \quad }c_{eff}^{i}=\frac{6}{\pi ^{2}}L\left( 
\frac{2}{3+\sqrt{5}}\right) =\frac{2}{5},
\end{equation}
respectively. As already mentioned the computation of the effective central
charge by means of (\ref{constTBA}) and (\ref{ceff}) is not always
rigorously justified. However, the explicit analytic computations of the full
scaling functions provided in the next section confirm these results. It is
intriguing to note that we recover in this way the effective central charges
for certain minimal models \cite{BPZ}. The case $\lambda =-1/2$ corresponds
to the minimal model ${\cal M}(3,5)$, whilst $\lambda =1$ corresponds to the
Yang-Lee model, i.e. ${\cal M}(2,5)$. Equation (\ref{ctba}) may also be
solved for different values of $\lambda $ and we may compute the effective
central charge by means of (\ref{ceff}). The dependence of $c_{eff}$ on $%
\lambda $ for the one-particle contribution is depicted in figure 1.
\footnote{The plot is presented only for $\lambda \geq -1$ since
eq.~(\ref{ctba}) possesses non-negative solutions only at
this interval. } 
This figure  suggests immediately that one may find other minimal models as
ultraviolet limits of (\ref{c(r)}) for different values of $\lambda $.
However, apart from $\lambda =0,-1,\infty $, the two values presented are
the only possible choices for a model with solely one particle leading to
accessible\footnote{%
In general a dilogarithmic identity is called accessible if it is of the
form $\sum_{i=1}^{N}L(y_{i})=k\pi ^{2}/6$, with $k$ being rational and $%
y_{i} $ algebraic \cite{Lewin}.} relations for dilogarithms. 
This implies that other 
rational $\lambda$ do not lead to rational values of $c_{eff}$. 

\section{Y-systems}

For some class of models it has turned out to be possible to carry out
certain manipulations on the TBA-equations such that the original integral
TBA-equations acquire the form of a set of functional equations in new
variables $Y_i$ (last ref. in \cite{TBAZam}). These functional equations have
the further virtue that unlike the original TBA-equations they do not
involve the mass spectrum and are commonly referred to as $Y$-systems. In
these new variables certain periodicities in the rapidities are exhibited
more clearly. These periodicities may then be utilized in order to express
the quantity $Y$ as a Fourier series, which in turn is useful to find
solution of the TBA-equations and expand the scaling function as a power
series in the scaling parameter. We will now demonstrate that similar
equations may be derived for a multi-particle system in which the dynamical
scattering is governed by the scattering matrix related to ADE-affine Toda
field theories and the statistical interaction is of general type.

We consider the modified version (in the sense of (\ref{SS})) of the
minimal part of the scattering matrix of ADE-affine Toda 
field theories. As was shown in 
\cite{Rava} these S-matrices lead to the identity 
\begin{equation}
\varphi _{ij}\left( \theta +\frac{i\pi }{h}\right) +\varphi _{ij}\left(
\theta -\frac{i\pi }{h}\right) =\sum\limits_{k=1}^{r}I_{ik}\varphi
_{kj}\left( \theta \right) -2\pi I_{ij}\delta \left( \theta \right) \quad ,
\label{RAD}
\end{equation}
where $h$ denotes the Coxeter number, $r$ the rank and $I$ the incidence
matrix of the Lie algebra. It is then straightforward to derive the ``$YZ-$%
system'' 
\begin{equation}
Y_{i}\left( \theta +\frac{i\pi }{h}\right) Y_{i}\left( \theta -\frac{i\pi }{h%
}\right) =\prod\limits_{j=1}^{r}Z_{j}\left( \theta \right) ^{I_{ij}}.
\label{YSYS}
\end{equation}
The quantities $Y\left( \theta \right) $ and $Z\left( \theta \right) $
involve now the statistical interaction in the form 
\begin{equation}
Y_{i}\left( \theta \right) =(1+x_{i}(\theta
))\prod\limits_{j=1}^{r}(1+x_{j}^{-1}(\theta ))^{-g_{ij}}\quad \text{and
\quad }\quad Z_{i}\left( \theta \right) =Y_{i}\left( \theta \right)
(1+x_{i}^{-1}(\theta ))\quad .  \label{Y}
\end{equation}
Equations (\ref{YSYS}) follow upon first adding (\ref{BFTBA}) at $\theta +%
\frac{i\pi }{h}$ and $\theta -\frac{i\pi }{h}$ and subtracting $I$ times (%
\ref{BFTBA}) at $\theta $ from the sum. Thereafter we employ the fact that
the masses of affine Toda field theory are proportional to the
Perron-Frobenius vector of the Cartan matrix, i.e. $%
\sum_{j=1}^{r}C_{ij}m_{j}=4\sin ^{2}(\pi /(2h))m_{i}$ \cite{Mass}. Then,
with the help of (\ref{RAD}) and (\ref{Y}), the equations (\ref{YSYS})
follow.

In comparison with (\ref{BFTBA}) the equations (\ref{YSYS}) have already the
virtue that they are simple functional equations and do not involve the mass
spectrum. However, in order to solve them we still have to express $Z$ in
terms of $Y$ or vice versa, which is not possible in general. However, once
the statistical interaction is specified this may be achieved. For instance,
for $g_{ij}=\delta _{ij}$ we obtain $Z_{i}=Y_{i}+1$ and recover the known
fermionic $Y$-system. In the bosonic case, i.e. $g_{ij}=0$ we obtain $%
Z_{i}=Y_{i}^{2}/(Y_{i}-1)$ and for ``semionic'' statistical interaction with $%
g_{ij}=\delta _{ij}/2$ we obtain $Z_{i}=\left( \sqrt{1+4Y_{i}^{2}}%
+1\right)^{2}/(4 Y_i) $.

\section{Examples}

\subsection{Ising model and Klein-Gordon Theory}

The most elementary examples which illustrate the features outlined above
more concretely is simply to consider the Ising model ($A_{1}$-minimal
affine Toda field theory) $S(\theta )=-1$ or the Klein-Gordon theory $%
S(\theta )=1.$ Then for both cases equation (\ref{BFTBA}) is solved
trivially. We obtain 
\begin{equation}
x(\theta )=\exp (rm\cosh \theta )-1\quad \text{and\quad }x(\theta )=\exp
(rm\cosh \theta )
\end{equation}
for bosonic- and fermionic statistical interaction, respectively. With the
help of these solutions we may then compute the entire scaling function. For
fermionic statistics we obtain from (\ref{c(r)})
\begin{equation}
c(r)=\frac{6}{\pi ^{2}}rm\sum\limits_{n=1}^{\infty }(-1)^{n+1}\frac{%
K_{1}(nrm)}{n}  \label{cf}
\end{equation}
whilst bosonic statistics yields 
\begin{equation}
c(r)=\frac{6}{\pi ^{2}}rm\sum\limits_{n=1}^{\infty }\frac{K_{1}(nrm)}{n}%
\quad ,  \label{cb}
\end{equation}
where $K_{1}(x)$ is a modified Bessel function. We depict these functions in
figure 3(b), referring to both of them by a slight abuse of notation 
as $A_1$ . One observes, that the difference in the statistical
interaction is effecting most severely the ultraviolet region. The two
scaling functions converge relatively fast towards each other in the
infrared regime. We shall encounter these feature also in other models.

Using the well-known property for the asymptotic behaviour of the modified
Bessel function $\lim\limits_{x\rightarrow 0}xK_{1}(x)=1$ leads to 
\begin{equation}
\lim\limits_{r\rightarrow 0}c(r)=-\frac{6}{\pi ^{2}}Li(-1)=\frac{1}{2}\quad 
\text{and\quad }\lim\limits_{r\rightarrow 0}c(r)=\frac{6}{\pi ^{2}}Li(1)=1
\end{equation}
for the fermionic and bosonic type equations, respectively. Here $Li(x)$ $%
=\sum_{n=1}^{\infty }\frac{x^{n}}{n^{2}}$ denotes Euler's dilogarithm \cite
{Lewin}.

We may also compute some less trivial cases. For instance taking the
statistical interaction to be of ``semionic'' type, i.e. $g=1/2,$ 
we obtain, after solving (\ref{BFTBA}), for the scaling function 
\begin{eqnarray}
c(r) &=&\frac{12r}{\pi ^{2}}m\int\limits_{0}^{\infty }d\theta \func{arsinh}%
\left[ \frac{1}{2}\exp (-rm\cosh \theta )\right] \cosh \theta \quad 
\nonumber \\
&=&\frac{12r}{\pi ^{2}}m\sum\limits_{n=0}^{\infty }(-1)^{n}\frac{(2n-1)!!}{%
(2n)!!(2n+1)2^{2n+1}}K_{1}((2n+1)rm)\quad .  \label{111}
\end{eqnarray}
Once more we carry out the ultraviolet limit with the help of the
asymptotics of the modified Bessel function 
\begin{equation}
c(0)=\frac{12}{\pi ^{2}}\sum\limits_{n=0}^{\infty }(-1)^{n}\frac{%
(2n-1)!!}{(2n)!!(2n+1)^{2}2^{2n+1}}=
\frac{12}{\pi^{2}} \int_0^{1/2} dt\, \frac{{\rm arsinh}\,t }{t}=
\frac{3}{5}\quad .
\end{equation}
This value of the effective central charge corresponds to the minimal model 
${\cal M}(3,5)$. This is of course what we expect from section 6, since with
the help of (\ref{trans}) it can be easily seen that we have thermodynamical
equivalence between the Calogero-Sutherland model with coupling $\lambda
=-1/2$ and fermionic statistical interaction and the Ising model with
``semionic'' statistical interaction.

For $g_{ij}=2\delta _{ij}$ we carry out a similar computation and obtain for
the scaling function 
\begin{equation}
c(r)=\frac{6r}{\pi ^{2}}m\int\limits_{0}^{\infty }d\theta \ln \left( \frac{1+%
\sqrt{1+4\exp (-rm\cosh \theta )}}{2}\right) \cosh \theta \quad .
\end{equation}
Once again we may perform the ultraviolet limit and obtain in this case 
$c_{eff}=2/5$, which corresponds to the minimal model ${\cal M}(2,5)$. 
This is in agreement with the results in section 6, since the 
Calogero-Sutherland model with coupling $\lambda =1$ and fermionic 
statistical interaction and the Ising model with statistical 
interaction $g_{ij}=2\delta _{ij}$ are thermodynamically equivalent.

\subsection{Scaling Potts and Yang-Lee Models}

Next we investigate the scaling Potts model in this context, which was
previously studied by Zamolodchikov \cite{TBAZam} with regard to
conventional fermionic statistics. The S-matrix of the scaling Potts model
equals the minimal S-matrix of $A_{2}$-affine Toda field theory \cite
{Roland2} and reads 
\begin{equation}
S_{11}(\theta )=S_{22}(\theta )=\frac{\sinh \left( \frac{\theta }{2}+\frac{%
i\pi }{3}\right) }{\sinh \left( \frac{\theta }{2}-\frac{i\pi }{3}\right) }%
\quad \text{and\quad }S_{12}(\theta )=-\frac{\sinh \left( \frac{\theta }{2}+%
\frac{i\pi }{6}\right) }{\sinh \left( \frac{\theta }{2}-\frac{i\pi }{6}%
\right) }\quad .  \label{A2}
\end{equation}
As commented in section 4, the S-matrix does not satisfy the bootstrap at $%
\theta =0.$ From our point of view it seems therefore more natural to use
its modification in the sense of (\ref{SS}) and employ bosonic statistics,
which of course by the equivalence principle leads to the same
TBA-equations. The two particles in the model are conjugate to each other,
i.e. $1=\bar{2}$, and consequently the masses are the same $m_{1}=m_{2}=m.$
The conjugate particle occurs as a bound state when two particles of the
same species scatter, for instance 1+1$\rightarrow $2. For the TBA-equation
we need 
\begin{equation}
\varphi _{11}(\theta )=\varphi _{22}(\theta )=\frac{-\sqrt{3}}{2\cosh \theta
+1}\quad \text{and\quad }\varphi _{12}(\theta )=\frac{\sqrt{3}}{1-2\cosh
\theta }\quad .
\end{equation}
Then for the fermionic statistics, equation (\ref{BFTBA}) can be solved
iteratively

\begin{equation}
\ln \left( x^{(n+1)}(\theta )\right) =rm\cosh \theta +\frac{2\sqrt{3}}{\pi }%
\int\limits_{-\infty }^{\infty }d\theta ^{\prime }\frac{\cosh (\theta
-\theta ^{\prime })}{1+2 \cosh 2(\theta -\theta ^{\prime })}\ln 
\left( 1+\frac{1}{x^{(n)}(\theta ^{\prime })}\right) .  \label{a2}
\end{equation}
Here we assumed that the $Z_{2}$-symmetry of the model will be preserved
such that $x_{1}(\theta )=x_{2}(\theta )$. Once more for $x(\theta )=\exp
(\varepsilon (\theta ))$ we recover the well-known TBA-equation of fermionic
type (first ref. in \cite{TBAZam}). It appears to be impossible to find
analytic solutions to this equation, but it is straightforward to solve it
numerically. Taking $\ln \left( x^{(0)}(\theta )\right) =rm\cosh \theta $
one can iterate this equation as indicated by the superscripts. Depending on
the value of $mr$ and $\theta $, convergence is achieved relatively quickly
(typically n\TEXTsymbol{<}50). The result is shown in figure 2(a) and
appears to be in complete agreement with the calculation in \cite{TBAZam}.
To make contact with the literature we introduced the quantity $L(\theta
)=\ln \left( 1+x^{-1}(\theta )\right) .$ One observes the typical behaviour
of $\lim_{mr\rightarrow 0} L(\theta )=const$ for some region of
$\theta$, which is required to derive (\ref{constTBA}).

We now consider a model with fermionic statistical interaction which
involves an S-matrix related to the scaling Potts model in the sense of (\ref
{anS}). According to the discussion in sections 4 and 5, this is equivalent
to considering the scaling Potts model in which the statistical interaction
is taken to be $g_{ij}=\delta _{ij}-N_{ij}$. Since 
\begin{equation}
N=\Delta ^{-}-\Delta ^{+}=-\frac{1}{2\pi }\int\limits_{-\infty }^{\infty
}d\theta \varphi (\theta )=C_{A_{2}}^{-1}\cdot I_{A_{2}} =\left( 
\begin{array}{ll}
\frac{1}{3} & \frac{2}{3} \\ 
\frac{2}{3} & \frac{1}{3}
\end{array}
\right) ,
\end{equation}
($C_{A_{2}}$ and $I_{A_{2}}$ denote here the Cartan matrix and
the incidence matrix of the $A_{2}$-Lie algebra), we obtain
\begin{equation}
 g = \left( \!\!
\begin{array}{rr}
\frac{2}{3} & -\frac{2}{3} \\ 
-\frac{2}{3} & \frac{2}{3}
\end{array}
\right) , \label{ndg}
\end{equation}
and the generalized TBA-equation (\ref{BFTBA}) becomes 
\begin{equation}
\ln \left( x(\theta )\right) =rm\cosh \theta -\ln \left( 1+x^{-1}(\theta
)\right) +\frac{2\sqrt{3}}{\pi }\int\limits_{-\infty }^{+\infty }d\theta
^{\prime }\frac{\cosh (\theta -\theta ^{\prime })\ln \left( 1+x^{-1}(\theta
^{\prime })\right) }{1+2 \cosh 2(\theta -\theta ^{\prime })}.
\label{53}
\end{equation}
At first sight $g_{ij}$ in (\ref{ndg}) seems to be inappropriate since
its off-diagonal elements are negative. 
However, due to the $Z_{2}$-symmetry this does not
pose a problem (c.f. section 2), since $g_{11}+g_{12}=0$ and 
$x_{1}(\theta)=x_{2}(\theta )$. In fact, this system may be 
thought of as a bosonic system with one particle species.
Equation  (\ref{53}) may be solved easily numerically in a
similar fashion as (\ref{a2}), whereas in comparison with (\ref{a2})
convergence is now achieved much faster (typically $n<20$). The result is
shown in figure 2(b).

Having solved the generalized TBA-equation for $x(\theta )$ we may also
compute the entire scaling function. The result of the numerical computation
is depicted in figure 3(b) for the two types of statistical interactions
presented in this subsection.
Notice that the conformal limit for the exotic statistics 
corresponds to $c=1$, which is in 
agreement with (\ref{37}). Hence in this limit we obtain a one
particle bosonic system.

Due to the fact that the scattering matrices of the scaling Potts model 
and the scaling Yang-Lee model are related as $S^{YL}(\theta)= 
S_{11}^{A_{2}}(\theta )S_{12}^{A_{2}}(\theta )$, the TBA-equations for the
two models are identical. The only difference occurs due the fact that in the
scaling Yang-Lee model there is only one instead of two particles present
and therefore the scaling functions equal each other up to a factor 2.

We finish the discussion of the  anyonic scaling Potts model with a brief
remark on other possible choices of exotic statistics. For example, we may
take the S-matrix of the scaling Potts model and choose the statistical
interaction to be of form $g_{ij}= g \delta_{ij} $. For this choice of
$g_{ij}$ the $Z_2$-symmetry is also present, so that
$x_{1}(\theta)=x_{2}(\theta )$. This leads, in particular, to a
simplification of the system (\ref{constTBA}). In fact, it reduces
to two equations of type (\ref{ctba}) with $i=1,2$ and 
$\lambda_1=\lambda_2=g$. Therefore, the effective central charge
describing conformal limit of this anyonic version of the scaling 
Potts model is given by $c_{SP}(g)=2c_{CS}(\lambda=g)$, 
where $c_{CS}(\lambda)$ is the effective central charge of the 
Calogero-Sutherland type model discussed above (see figure 1). 
This also means that the  conformal limit of the scaling Yang-Lee model with 
statistical interaction $g$ is described by $c_{YL}(g)=c_{CS}(\lambda=g)$ 
(of course, as we discussed in section 6, this does not imply global
equivalence with the Calogero-Sutherland model).

\subsection{$A_{3}$-minimal affine Toda Theory}

Finally we present the fermionic computation for the easiest model, which
contains at least two different mass values , i.e. $A_{3}$-affine Toda field
theory. The minimal S-matrix reads \cite{Roland2} 
\begin{equation}
\begin{array}{ll}
S_{11}(\theta )=S_{33}(\theta )=\frac{\sinh \left( \frac{\theta }{2}+\frac{%
i\pi }{4}\right) }{\sinh \left( \frac{\theta }{2}-\frac{i\pi }{4}\right) } & 
S_{13}(\theta )=-\frac{\sinh \left( \frac{\theta }{2}+\frac{i\pi }{4}\right) 
}{\sinh \left( \frac{\theta }{2}-\frac{i\pi }{4}\right) } \\ 
S_{12}(\theta )=S_{23}(\theta )=\frac{\sinh \left( \frac{\theta }{2}+\frac{%
i\pi }{8}\right) }{\sinh \left( \frac{\theta }{2}-\frac{i\pi }{8}\right) }%
\frac{\sinh \left( \frac{\theta }{2}+\frac{3\pi i}{8}\right) }{\sinh \left( 
\frac{\theta }{2}-\frac{3\pi i}{8}\right) } & S_{22}(\theta )=-\left( \frac{%
\sinh \left( \frac{\theta }{2}+\frac{i\pi }{4}\right) }{\sinh \left( \frac{%
\theta }{2}-\frac{i\pi }{4}\right) }\right) ^{2}\,.
\end{array}
\end{equation}
Hence 
\begin{eqnarray}
\varphi _{11}(\theta ) &=&\varphi _{33}(\theta )=\varphi _{13}(\theta )=%
\frac{1}{2}\varphi _{22}(\theta )=-\frac{1}{\cosh \theta } \\
\varphi _{12}(\theta ) &=&\varphi _{23}(\theta )=-\frac{2\sqrt{2}\cosh
\theta }{\cosh 2\theta }\quad .
\end{eqnarray}
The masses are given by $m_{1}=m_{3}=m/\sqrt{2}$ and $m_{3}=m.$ Again we
assume that the $Z_{2}$-symmetry is preserved, such that $x_{1}(\theta
)=x_{3}(\theta )$. The TBA-equations for fermionic statistics read 
\begin{eqnarray*}
\ln \left( x_{1}(\theta )\right) &=&\frac{rm}{\sqrt{2}}\cosh \theta
+\int\limits_{-\infty }^{\infty }d\theta ^{\prime }\left( \frac{L_{1}(\theta
^{\prime })}{\pi \cosh (\theta -\theta ^{\prime })}+\frac{\sqrt{2}\cosh
(\theta -\theta ^{\prime })L_{2}(\theta ^{\prime })}{\pi \cosh 2(\theta
-\theta ^{\prime })}\right) , \\
\ln \left( x_{2}(\theta )\right) &=&rm\cosh \theta +\int\limits_{-\infty
}^{\infty }d\theta ^{\prime }\left( \frac{L_{2}(\theta ^{\prime })}{\pi
\cosh (\theta -\theta ^{\prime })}+\frac{\sqrt{8}\cosh (\theta -\theta
^{\prime })L_{1}(\theta ^{\prime })}{\pi \cosh 2(\theta -\theta ^{\prime })}%
\right) ,
\end{eqnarray*}
where $L_{i}(\theta )=\ln \left( 1+x_{i}^{-1}(\theta )\right) $ for $i=1,2$.
Once again we may solve these equations numerically and compute the entire
scaling function. The result is depicted in figure 3(b). The functions $%
L_{i}(\theta )$ exhibit the typical plateau as $mr$ approaches zero. In
comparison with the other models we observe that scaling functions for
systems with the same statistical interaction have qualitatively the same
shape. Similarly as in the previous subsection we could now also consider
the anyonic system with $g_{ij}=\delta _{ij}-N_{ij}.$ However, in this case
we would have an overall negative contribution of $\rho _{1}^{r}$ to the
density $\rho _{2}$ in (\ref{Dichte}), which as discussed in section 2 seems
inappropriate from a physical point of view. We also observed that in the
numerical computations singularities occur. However, we may perfectly well
choose a different type of statistics which leads to  satisfactory equations. 
For instance, we may choose  $g_{ij}= g \delta_{ij} $. 
In this case equations (\ref{constTBA}) and (\ref{ceff}) predict that
for $g=0,\, 1/2,\, 1,\, 2$ the conformal limit is described by
$c_{eff}$ taking the following values: 
$c_{eff}(0)\approx 1.16$, $c_{eff}(1/2)\approx 1.07$,
$c_{eff}(1) = 1$ and $c_{eff}(2)\approx 0.895$.
One may conjecture that $c_{eff}(g)$ is a monotonically decreasing
function. 
A detailed investigation of whole scaling functions for admittable
choices of statistical interaction is left for future studies.

\section{Conclusion}

For a multi-particle system which involves a factorizable scattering matrix
describing the dynamical interaction and a statistical interaction governed
by Haldane statistics we derived the TBA-equations. These equations may be
solved by the same means as the conventional TBA-equations of fermionic and
bosonic type and allow the computation of the entire scaling function.

The behaviour of the scaling functions depicted in figure 3(b) suggests the
validity of the conjecture \cite{TBAZam}, that the series for the scaling
function in the scaling parameter $r$ commences with a constant and
thereafter involves quadratic and higher powers in $r,$ for fermionic
type of statistics. For exotic statistics these features don't seem to be
evident.

The question of how to select a particular statistics without the prior
knowledge of the nature of the particles remains to be clarified.

It would be very interesting  to generalize these kind of considerations
also to the situation in which the dynamical scattering is described
by non-diagonal S-matrices (e.g. affine Toda field theory with purely
imaginary coupling constant), to the excited TBA-equations
\cite{TBAZam,TBA} or even more exotic statistics \cite{IIG}.

After the completion of our manuscript Dr. Ilinski pointed out to us,
that there exists an ongoing dispute \cite{Il2} about how to achieve
compatibility between equations (1) and (2), that is to provide a
prescription for the counting of the states involving (2) which would lead
to (1). There are even doubts whether it is at all possible to achieve
compatibility between the two equations. What our manuscript concerns, this
is not a crucial issue, since we could also start with a much weaker 
assumption. We need only (1) with $d_i$ replaced according to (2).

We are also grateful to Dr. Hikami for pointing out reference \cite{Hik}
to us, in which one also finds a discussion on ideal $g$-on gas systems
with fractional exclusion statistics. In particular our figure 1 is
similar to the figure in there.

\newpage

{\bf Acknowledgment:} We would like to thank C. Ahn, H. Babujian, C.
Figueira de Morisson Faria, M. Karowski and R. K\"{o}berle for useful
discussions and comments. A.B. is grateful to the Volkswagen Stiftung and
Russian Fund for Fundamental Investigations for financial support and
to the members of the Institut f\"{u}r Theoretische Physik for their  
hospitality. A.F. is grateful to the Deutsche
Forschungsgemeinschaft (Sfb288) for support.


\end{document}